\documentclass[aps,prb,superscriptaddress,twocolumn]{revtex4-2}

\usepackage{graphics}
\usepackage{graphicx}% Include figure files
\usepackage{dcolumn}% Align table columns on decimal point
\usepackage{bm}% bold math
\usepackage[usenames]{color}
\usepackage[ %showframe,%Uncomment any one of the following lines to test
margin=0.75in,
]{geometry}
\usepackage[normalem]{ulem}
\usepackage{amsmath}

\graphicspath{{./figs/}}

\def\Ho227{Ho$_2$Ti$_2$O$_7$}
\def\Dy215{Dy$_2$TiO$_5$}
\def\Tb227{Tb$_2$Sn$_2$O$_7$}
\def\An{$\mathrm{\AA}$}

\begin{document}

\title{Magnetic Pair Distribution Function and Half Polarized Neutron Powder Diffraction at the HB-2A Powder Diffractometer}

	\author{Raju Baral}
        \email{baralr@ornl.gov}
	\affiliation{ %
		Neutron Scattering Division, Oak Ridge National Laboratory, Oak Ridge, Tennessee 37831, USA.
	} %
	
\author{Danielle R.~Yahne}
	\affiliation{ %
		Neutron Scattering Division, Oak Ridge National Laboratory, Oak Ridge, Tennessee 37831, USA.
	} %

\author{Malcolm J.~Cochran}
	\affiliation{ %
		Neutron Scattering Division, Oak Ridge National Laboratory, Oak Ridge, Tennessee 37831, USA.
	} %

\author{Matthew~Powell}
\affiliation{Department of Chemistry and Center for Optical Materials Science and Engineering Technologies, Clemson University, Clemson, South Carolina 29634, USA}	

\author{Joseph~W.~Kolis}
\affiliation{Department of Chemistry and Center for Optical Materials Science and Engineering Technologies, Clemson University, Clemson, South Carolina 29634, USA}	

\author{Haidong Zhou}
\affiliation{Department of Physics and Astronomy, University of Tennessee, Knoxville, Tennessee 37996, USA.}

	\author{Stuart Calder}
        \email{caldersa@ornl.gov}
	\affiliation{ %
		Neutron Scattering Division, Oak Ridge National Laboratory, Oak Ridge, Tennessee 37831, USA.
	} %
	
\begin{abstract}
Local magnetic order and anisotropy are often central for understanding fundamental behavior and emergent functional properties in quantum materials and beyond. Advances in neutron powder diffraction experiments and analysis tools now allow for quantitative determination. We demonstrate this here with complementary total neutron scattering and polarized neutron measurements on the HB-2A neutron powder diffractometer at the High Flux Isotope Reactor (HFIR). In recent years, magnetic pair distribution function (mPDF) analysis has emerged as a powerful technique for probing local magnetic spin ordering of magnetic materials. This method can be broadly applied to any magnetic material but is particularly effective for studying systems with short-range magnetic order, such as materials with reduced dimensionality, geometrically frustrated magnets, thermoelectrics, multiferroics, and correlated paramagnets. Magnetic anisotropy often underpins the short-range order adopted. Half-polarized neutron powder diffraction (pNPD) can be used to determine the local susceptibility tensor on the magnetic sites to quantify the magnetic anisotropy. Combining the techniques of mPDF and pNPD can therefore provide valuable insights into local magnetic behavior. A series of measurements optimized for these techniques are presented as exemplar cases focused on frustrated materials where short-range order dominates, these include measurements to ultra-low temperature ($<100$ mK) not typically accessible for such experiments.  

\end{abstract}
	
\maketitle

\section{Introduction}	

The fundamental behavior of materials depends on the structural arrangement of the constituent atoms. Many materials can be considered crystalline and the average long-range structure provides a robust framework, however, when well-defined local order or competing interactions are present the behavior can be profoundly altered from the average. Growing evidence has highlighted that magnetic and structural short-range order can play a significant role in determining various functional properties of materials, with examples that include the paramagnon drag effect and giant spontaneous magnetostriction in the altermagnet MnTe \cite{baral2022realspace, baral2023giant,xu2017performance}, negative thermal expansion in ZrW$_2$O$_8$ \cite{tucker2017ZrW2O8}, thermoelectric efficiency \cite{xu2017performance}, advancing magnetic resonance imaging (MRI) technology \cite{na2009inorganic} and controlling the efficiency of Li-ion batteries \cite{wang2017}. Such magnetic and structural short-range order can be induced through several routes, including chemical disorder, external perturbations, geometric frustration of spins or charge, reduced dimensionality, and is present to varying degrees in most, if not all, materials. Hence, a comprehensive understanding of both short-range and long-range magnetic correlations is essential for fundamental understanding and unlocking the full potential of functional materials.   

Characterizing local order is non-trivial, but there is strong motivation to develop tools to probe short-range order. Neutron powder diffraction provides an extremely powerful technique to quantitatively characterize the long-range crystalline and magnetic structures in straightforward measurements. Neutron diffraction has been central to the development of understanding long-range magnetic structures since the inception of the technique \cite{PhysRev.76.1256.2, PhysRev.83.333}. Furthermore, neutron scattering can be extended to investigate not only long-range order in the form of Bragg scattering but also short-range order through consideration of the whole scattering pattern with the neutron total scattering technique. For atomic scattering, this is now an established tool and has been extensively used to study the local atomic structure of complex and disordered materials \cite{egami2012underneath}. The total scattering technique involves performing a Fourier transform on the full diffraction pattern to convert from momentum space ($Q$-space) into real space, providing a direct and intuitive view of atom-atom distributions at various distances through the pair distribution function (PDF). Recently, this atomic PDF concept was extended to magnetism to develop the magnetic pair distribution function (mPDF) technique to study short-range spin-spin correlations directly in real space using neutron total scattering experiments~\cite{Frandsen2014, frandsen2015mPDFMnO}. The mPDF method involves the Fourier transformation of scattering from magnetic moments, thus providing real-space information on local magnetic order \cite{frandsen2015mPDFMnO,frandsen2016verification,frandsen2022magnetic}. The technique of mPDF is still in its infancy, however, there are a growing number of examples of successful mPDF experiments that have been conducted on a growing number of neutron instruments \cite{proffen2002building,neuefeind2012nanoscale,frandsen2022magnetic,kodama2017alternative,yang2024unveiling,lefranccois2019spin,dun2021neutron, baral2024localSpinStructure,baral2025_MST_HB2A,hatt2024_ZnMnTe_HB2A}. 

The specific magnetic moment direction adopted in the material of interest is governed by magnetic anisotropy, with underlying factors being the surrounding crystalline field environment and spin-orbit coupling. A parameter to quantify this is the susceptibility, which is a measure of the magnetization in an applied field. When magnetic anisotropy is present, the magnetic susceptibility is described by a second-rank tensor, rather than a simple scalar. The susceptibility tensor can be transformed into magnetic ellipsoids in the crystallographic frame to visualize elongation, flattening, or in the absence of local anisotropy, these ellipsoids transform into spheres, with their radii proportional to the induced magnetization. Determining the susceptibility tensor provides a direct route to quantify the magnetic anisotropy, which informs the short- and long-range orders adopted in materials. Once again, extending traditional neutron powder diffraction allows for this to be determined. Specifically, using half-polarized neutron powder diffraction (pNPD) provides the required sensitivity. Although this technique was developed and primarily only applied to single crystal neutron measurements \cite{AGukasov_2002}, it has recently been shown that powder measurements can provide equivalent information \cite{Gukasov_2010, PhysRevResearch.1.033100}. pNPD is experimentally the simplest polarized technique, with only two measurements required of spin-up and spin-down incident neutron polarization. This, however, allows access to further scattering terms that result in enhanced sensitivity to access typically inaccessible moments of  under $0.1\mu_{\rm B}$ in powder samples. pNPD requires the sample to have a net moment, so measurements are typically performed in the paramagnetic regime under an applied field. Consequently, pNPD works equally well for materials with short- or long-range order, since measurements are not required to be performed below any ordering temperature. pNPD has been applied to areas such as anisotropy in molecular-based magnets \cite{magnetochemistry7120158, Klahn:lt5038}, coordination polymers \cite{PhysRevB.86.054431}, magnetic metal-organic frameworks \cite{Calder:cam5003, PhysRevMaterials.7.124408, PhysRevMaterials.6.124407, doi:10.1021/jacsau.2c00575}, 5$d$ based magnets \cite{PhysRevB.93.134426}, magnetic nanoparticles \cite{PhysRevB.90.134427, https://doi.org/10.1002/smtd.202201725}, magnetocaloric materials \cite{TENER2021167827}. 

An area where short-range magnetic order and anisotropy dominate is geometrically frustrated magnets, particularly those containing rare-earth magnetic ions. These materials have well-ordered crystalline structures however the magnetic ion resides on a lattice where long-range spin-ordering is inhibited due to the geometry, which typically consists of triangular units in two or three dimensions \cite{greedan2001}. The magnetic frustration leads to a variety of spin states, with emergent behavior that hinges on the equilibrium between magnetic exchange interactions, dipolar interactions, and crystal field effects. In general, geometrically frustrated magnets have formed a well-spring of materials that can host exotic phenomena. In particular, geometrically frustrated rare-earth pyrochlores, $R_2$$M_2$O$_7$ ($R$ = rare earth, $M$ = transition metal) have formed an enduring interest in condensed matter research due to the lattice topology of a highly frustrated corner-sharing tetrahedra of magnetic ions. A wide variety of physical phenomena have been predicted and observed in pyrochlores, including classic and quantum spin ice with emergent magnetic monopoles, spin liquid candidates, order by disorder, multipolar ordering, the anomalous Hall effect and metal-insulator transitions \cite{RevModPhys.82.53, doi:10.1146/annurev-conmatphys-022317-110520}. 

Here, we demonstrate how neutron powder diffraction can be extended to perform total neutron scattering mPDF measurements and polarized neutron scattering through pNPD measurements to gain a comprehensive understanding of both long-range and short-range magnetic order. These two distinct measurements are performed on a single instrument, HB-2A at HFIR. This instrument has traditionally been utilized for long-range magnetic structure determination and has ultra-low temperature and magnetic fields available, with the option for multi-sample changers, that match the requirements for mPDF and pNPD. HB-2A offers unique sample conditions not typically found on traditional PDF instruments, in particular the ability to access ultra-low temperatures, which is often required for quantum material research. In addition HB-2A is the only dedicated neutron powder diffraction instrument in the United States where pNPD measurements can be performed. The materials chosen are all rare-earth-based geometrically frustrated magnets, with the exception of MnO, which has become a de facto standard for mPDF. We stress, however, that these techniques, particularly when combined, can be utilized for investigations in a large variety of scientific fields, including magnetic anisotropy in organic materials and single molecular magnets, functional materials with properties such as thermoelectricity and magneto-caloric behavior, multiferroics, low-dimensional spintronic materials, magnetic nanoparticles, to name a few examples.  

\section{Experimental methods and formalisms}

\subsection{HB-2A Powder Diffractometer}

Neutron powder diffraction measurements were carried out on the HB-2A powder diffractometer at the High Flux Isotope Reactor (HFIR), Oak Ridge National Laboratory (ORNL) \cite{garlea2010HFIR, powderHB2A2018}. HB-2A is a constant wavelength thermal diffractometer with a balance of high flux and high resolution. Traditionally, this instrument has been utilized for studies of long-range magnetic structures under variable temperature, magnetic field or pressure, however, it has the versatility to be expanded to total scattering studies and polarized measurements. The wavelength is selected from a germanium monochromator with a 90$^{\circ}$ take-off angle using three principle reflections: Ge113 (2.41 \AA), Ge115 (1.54 \AA), Ge117 (1.12 \AA). The pre-mono collimation is open, and the pre-detector collimation consists of $12^{\prime}$ Soller collimators, with the flexibility of the pre-sample collimation between open or $21^{\prime}$, with other options of $12^{\prime}$, and $31^{\prime}$ available. The current detector array consists of 44 $^3$He detectors spaced by 2.8$^{\circ}$  that have to be stepped in 0.05$^{\circ}$ steps to fill in the gaps and give full 120$^{\circ}$ coverage within the range 5$^{\circ}$ to 155$^{\circ}$. A planned detector upgrade to implement continuous detector coverage, termed MIDAS, will greatly improve counting efficiency by an order of magnitude.

HB-2A has developed unique sample changers that make use of the versatile sample stage with two rotations and x,y,z translation. Consequently, both internal motion sample changers and external motion sample changers are used, enabling multiple samples in dilution refrigerators (0.05 K-300 K, 3-samples), He$^3$ dry cryostats (0.3 K-300 K, 14-samples), He$^4$ cryostats (1.5K – 300 K, 23-samples), closed cycle refrigerators (4 K – 300 K, 6-samples) and cryomagnets (1.5 K-300 K, 0-6 T, 3-samples). This wide range of sample changers has greatly enhanced the ability to perform the total scattering measurements discussed here by enabling routine collection of empty sample holders and background measurements in the same conditions as the sample of interest, even at ultra-low temperatures.  

\subsection{Magnetic PDF measurements and Data reduction}
\label{mPDFsection}

\subsubsection{mPDF formalism}

The magnetic PDF ($g_{\mathrm{mag}}(r)$) for a powder sample having identical localized spins is given by \cite{Frandsen2014,frandsen2015mPDFMnO}

\begin{align}
g_{\text{mag}}(r) &= \frac{2}{\pi} \int_{Q_{\text{min}}}^{\infty} Q
\Bigg[  \frac{\left( \frac{d\sigma}{d\Omega} \right)_{\text{mag}}}
{\frac{2}{3} N_s S(S+1)(\gamma r_0)^2 [f_m(Q)]^2}-1 \Bigg]  \nonumber \\
&\quad \times \sin(Qr) \, dQ, \label{eq:mpdf1} \\
&=  \frac{3}{2S(S+1)N_s} \Bigg[
\sum_{i \neq j} \left( \frac{A_{ij}}{r} \, \delta(r - r_{ij}) \right) \nonumber \\
&\quad + B_{ij} \, \frac{r}{r_{ij}^3} \, \Theta(r_{ij} - r)  
- 4 \pi \rho r \cdot \frac{2}{3} m^{2}
\Bigg]. \label{eq:mpdf2}
\end{align}

Equation \ref{eq:mpdf1} and \ref{eq:mpdf2} define the experimental and theoretical formulations of mPDF for a given magnetic structure. One immediate consequence to note is that the mPDF function involves the delta function at spin-pair separation distances and is modulated by the orientational term $A_{ij} = \langle S_i^y S_j^y \rangle$, which is positive for ferromagnetic alignment along the $y$-direction and negative for antiferromagnetic alignment. Here $S_i$ and $S_j$ represent individual spins separated by a distance $r_{ij}$. This provides a powerful model-independent way to view spin directions in real space, even before any analysis is undertaken.

Considering the equations further, $Q$ represents the magnitude of the scattering vector while $Q_{\mathrm{min}}$ is the minimum measured scattering vector. The term $(\frac{d\sigma}{d\Omega})_{\mathrm{mag}}$ denotes the magnetic differential scattering cross-section. Here, $N_s$ is the number of spins in the system, $S$ represents the spin quantum number in units of $\hbar$, $r$ is the real space distance. The parameter $\gamma$ represents the neutron magnetic moment in units of nuclear magnetons, $r_0$ is the classical electron radius, $f_m(Q)$ is the magnetic form factor, $\Theta(x)$ represents the Heaviside step function, $\rho$, and $m$ are the spin density and average magnetic moment in Bohr magnetons which is zero for antiferromagnetic material.  Additionally, $B_{ij}=2\langle S_i^x S_j^x \rangle - \langle S_i^y S_j^y \rangle$. mPDF also contains a term that is linear in $r$, which is absent in atomic PDF, and arises due to the fact that magnetic neutron scattering depends on spatial as well as orientational correlations of magnetic moments. 

When the PDF is generated using standard PDF protocols or with software such as PDFgetN3 \cite{Juhas2018PDFgetN3}, it is obtained from the Fourier transform of the total scattering intensity as:
\begin{eqnarray*}
G_{\text{tot}}(r) &=& \mathcal{F} \left\{ Q \left( \frac{I_{\text{tot}}}{N_a \langle b \rangle^2} - \frac{\langle b^2 \rangle}{\langle b \rangle^2} \right) \right\}
\end{eqnarray*}

Expanding the total scattered intensity into nuclear and magnetic components:
\begin{eqnarray*}
G_{\text{tot}}(r) &=& \mathcal{F} \left\{ Q \left( \frac{I_{\text{N}}}{N_a \langle b \rangle^2} - \frac{\langle b^2 \rangle}{\langle b \rangle^2} \right) \right\} + \mathcal{F} \left\{ Q \left( \frac{I_{\text{M}}}{N_a \langle b \rangle^2}\right) \right\}
\end{eqnarray*}

where $I_{\mathrm{tot}}$ is the total scattered intensity, consisting of the nuclear intensity ($I_N$) and magnetic scattered intensity ($I_M$). $N_a$ represents the number of atoms while $b$ denotes the nuclear scattering length. The term $\langle b \rangle$ is the average scattering length over all nuclei present in the sample. Thus, the total PDF can be expressed as:

\begin{eqnarray*}
G_{\text{tot}}(r) &=& G_{a}(r) + \frac{d_{\text{mag}}(r)}{N_a \langle b \rangle^2}
\end{eqnarray*}

where $G_{a}(r)$ is the atomic PDF and $d_{\mathrm{mag}}(r) = \mathcal{F} \left\{ Q \operatorname{I_m}(Q) \right\}$
 is termed the \textit{non-deconvoluted} or \textit{unnormalized}  mPDF since it is not divided by the magnetic form factor. Using the convolution theorem, $d_{\mathrm{mag}}(r)$ can be expressed as:

\begin{eqnarray*}
    d_{\text{mag}}(r) &=& C_1 \times g_{\text{mag}} * S(r) + C_2 \times \frac{dS}{dr}
\end{eqnarray*}
In the fully ordered state, the constants $C_1$ and $C_2$ are related by $\frac{C_1}{C_2} = -\frac{1}{\sqrt{2\pi}}$, and $S(r) = \mathcal{F} \{ f_{m}(Q) \} * \mathcal{F} \{ f_m(Q) \}$.

The primary distinction between the \textit{non-deconvoluted} mPDF ($d_{\mathrm{mag}}$) and \textit{deconvoluted} mPDF ($g_{\mathrm{mag}}$) patterns lies in the treatment of the magnetic form factor. In the \textit{deconvoluted} mPDF ($g_{\mathrm{mag}}$) or \textit{normalized} mPDF pattern, the total scattering intensity is divided by the squared magnetic form factor before applying the Fourier transform. A consequence is the \textit{normalized} mPDF pattern has better resolution in real space compared to the \textit{unnormalized} pattern.

\subsubsection{mPDF experiments}

Neutron total scattering measurements can be performed on the HB-2A beamline in a similar manner to traditional Bragg scattering studies. A key distinction is the requirement to remove background scattering from the sample holder and instrument during the data reduction. This necessitates the collection of empty sample holder measurements and empty instrument measurements. However, in the case of performing a direct temperature subtraction, this step can often be removed, as we will consider below. The three wavelength configurations on HB-2A allow flexibility in optimizing the experimental setup based on the specific experimental requirements. The shortest wavelength of $1.12~\mathrm{\AA}$ enables the largest $Q$-range of $10.8~\mathrm{\AA^{-1}}$, while the $1.54~\mathrm{\AA}$ wavelength provides a $Q$-range of $7.9~\mathrm{\AA^{-1}}$. The longest wavelength of $2.41~\mathrm{\AA}$ has a Q-range of only $5.0~\mathrm{\AA^{-1}}$ and is better suited to reciprocal space studies of long-range order or diffuse scattering rather than total scattering \cite{paddison2013spinvert}. For mPDF analysis, the choice between $1.12~\mathrm{\AA}$ and $1.54~\mathrm{\AA}$ wavelengths depends on the relative strength of the magnetic signal compared to the atomic PDF signal of the material, since this will determine the method for data reduction and analysis. The measurement temperatures of interest and the symmetry of the crystal structure also play a significant role in optimizing the choice of wavelength.
\subsubsection{mPDF Method I:}
\textit{Atomic and magnetic refinement in real space.}
The first method for mPDF analysis described is best suited to materials with a large magnetic signal compared to the structure and involves real-space modeling of both the atomic and magnetic signals. This method may also be best suited to measurements at high temperatures, above 100 K, where temperature subtraction to isolate the magnetic signal may introduce too many unwanted artifacts in the data. The wavelength of $1.12~\mathrm{\AA}$ provides a higher real-space resolution from the higher $Q_{\text{max}}$ ($Q_{\text{max}}=10.8~\mathrm{\AA}^{-1}$) for data collection using the HB-2A instrument and is best suited for this method. The instrumental background and empty-can scattering are required to be collected and subtracted from the data, which is then Fourier transformed with PDFgetN3~\cite{Juhas2018PDFgetN3} to generate the  PDF data.

The PDF data obtained contains the atomic as well as the magnetic signals and has to be modeled in real space. Typically the atomic scattering signal is significantly larger than the magnetic signal. Consequently, to isolate the magnetic signal and perform the magnetic refinement of the total neutron data, an initial atomic refinement has to be undertaken, which can be done using PDFgui or similar software. The fit residual of the atomic refinement is treated as the magnetic PDF data and this is used for the magnetic PDF refinement. We can also directly use a python package \texttt{diffpy.mpdf}~\cite{frandsen2022diffpympdf} to perform atomic and magnetic co-refinement. Details of atomic and magnetic co-refinements can be found on the Ref.\cite{FrandsenGroup_mPDFtutorial}. The mPDF pattern separated in real space from this method provides an \textit{unnormalized} pattern. This has a lower resolution; however, the magnetic signal is inherently broader than the atomic signal, and therefore this loss of resolution may not noticeably impact the analysis.

\subsubsection{mPDF Method II:}
\textit{Temperature subtraction in $Q$-space.}
The $Q$-space temperature subtraction approach is well suited for magnetic materials where all measurements will be conducted at low temperatures. Since the change in lattice constants will be small between the lowest and highest temperatures measured, this reduces any artifacts introduced from the temperature subtraction. Generally, all measurements should be below 75-100 K, however this will vary between materials.  

In this method, a high-temperature dataset is subtracted from low-temperature data in $Q$-space, effectively removing the nuclear contribution and isolating the magnetic signal. The reference high temperature must be selected such that the material is purely in a paramagnetic state; otherwise, the diffuse signal may be suppressed during the temperature subtraction. The residual magnetic signal obtained in reciprocal space is then Fourier-transformed to obtain the unnormalized mPDF pattern. If we divide the magnetic signal in reciprocal space by the square of the magnetic form factor before performing the Fourier transform, we obtain the normalized mPDF pattern in real space.
For this approach, using the longer wavelength of $1.54~\mathrm{\AA}$ may provide better resolution and better signal quality from the HB-2A instrument compared to $1.12~\mathrm{\AA}$ which has lower flux. In this approach, the magnetic scattering patterns can be Fourier transformed with $Q_{\text{max}}=7.9~\mathrm{\AA}^{-1}$ to generate the real-space mPDF patterns.

\subsection{Half-polarized neutron powder diffraction: Instrumentation and data analysis}

Half-polarized neutron powder diffraction (pNPD) requires only the polarization of the incident neutron beam and only two measurements with the neutron polarization in opposite directions: spin-up and spin-down. As such, pNPD is the simplest polarized technique to implement on a neutron scattering instrument, yet pNPD offers powerful enhancements to traditional unpolarized neutron powder diffraction. For the technique to work, the material of interest has to have a net magnetization, otherwise the sample will depolarize the beam. This necessitates the use of a magnetic field at the sample position. This sample magnetization can be field-induced in the paramagnetic temperature region in any material and suitable analysis provides a direct measure of the local site susceptibility tensor to reveal the local anisotropy of the magnetic ion. Alternatively, pNPD can be employed to measure very weak magnetic signals ($<$0.1$\mu_B$) in ordered ferromagnetic, ferrimagnetic or canted-antiferromagnetic materials below their transition temperature \cite{PhysRevResearch.1.033100}, even in this case an applied magnetic field is required to align domains and ensure a net magnetization and avoid beam depolarization. Here, we will only consider the case of field-induced magnetization in the paramagnetic phase to extract the local site susceptibility tensor of the magnetic ion. 

\subsubsection{Determining the local site susceptibility tensor}

The magnetic susceptibility ($\chi$) is a measure of the magnetization of a material in an applied magnetic field and is defined as $\chi=M/H$, where $M$ is the magnetization from the net magnetic moments and $H$ is the applied magnetic field. In most crystalline materials the influence of spin-orbit coupling induces anisotropy that results in $\chi$ not being a scalar quantity, but a second-rank tensor with the form: 

\begin{eqnarray}
\hat{\chi}_{ij}  = 
\begin{pmatrix}
{\chi_{11}} & {\chi_{12}} & {\chi_{13}}\\
{\chi_{12}} & {\chi_{22}} & {\chi_{23}}\\
{\chi_{13}} & {\chi_{23}} & {\chi_{33}}\\
\end{pmatrix}
=\frac{{\vec M_i}}{{\vec H}_j},
\label{sus_tensor}
\end{eqnarray}

where the site symmetry determines any constraints and the allowed non-zero tensor components $\chi_{ij}$. The $\chi_{ij}$ values can be extracted directly from pNPD measurements. This provides an insight into the local anisotropy of the magnetic ion. This method is applicable for materials in the linear M/H regime and does not include any further interactions, such as magnetic exchange interaction. 

%%%%%%%%%%%%
% Begin Figure
%%%%%%%%%%%%
%trim is left, bottom, right, top
\begin{figure}
    \centering
    \includegraphics[trim=0cm 0.0cm 0cm 0cm,clip=true,width=1.0\linewidth]{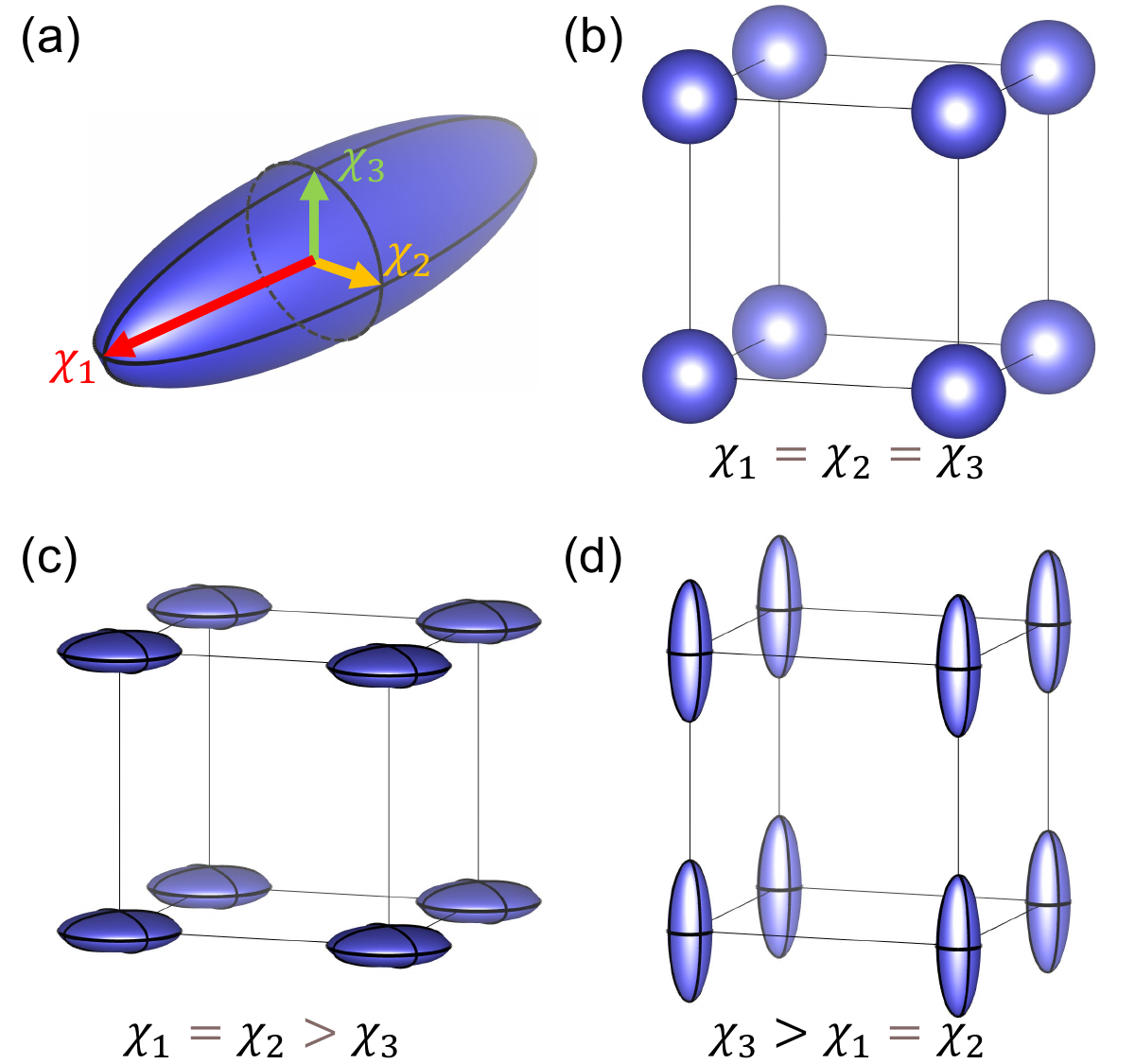}
    \caption{\label{fig:Ellipsoids} (a) Magnetization ellipsoids from the local site susceptibility tensor. (b) Isotropic Heisenberg spins, (c) XY-planar anisotropy, and (d) Ising spins.
    }
\end{figure}
%%%%%%%%%%%%
% End Figure
%%%%%%%%%%%%

%%%%%%%%%%%%
% Begin Figure
%%%%%%%%%%%%
%trim is left, bottom, right, top
\begin{figure*}
    \centering
    \includegraphics[trim=0cm 1.0cm 0cm 1.0cm,clip=true,width=0.99\linewidth]{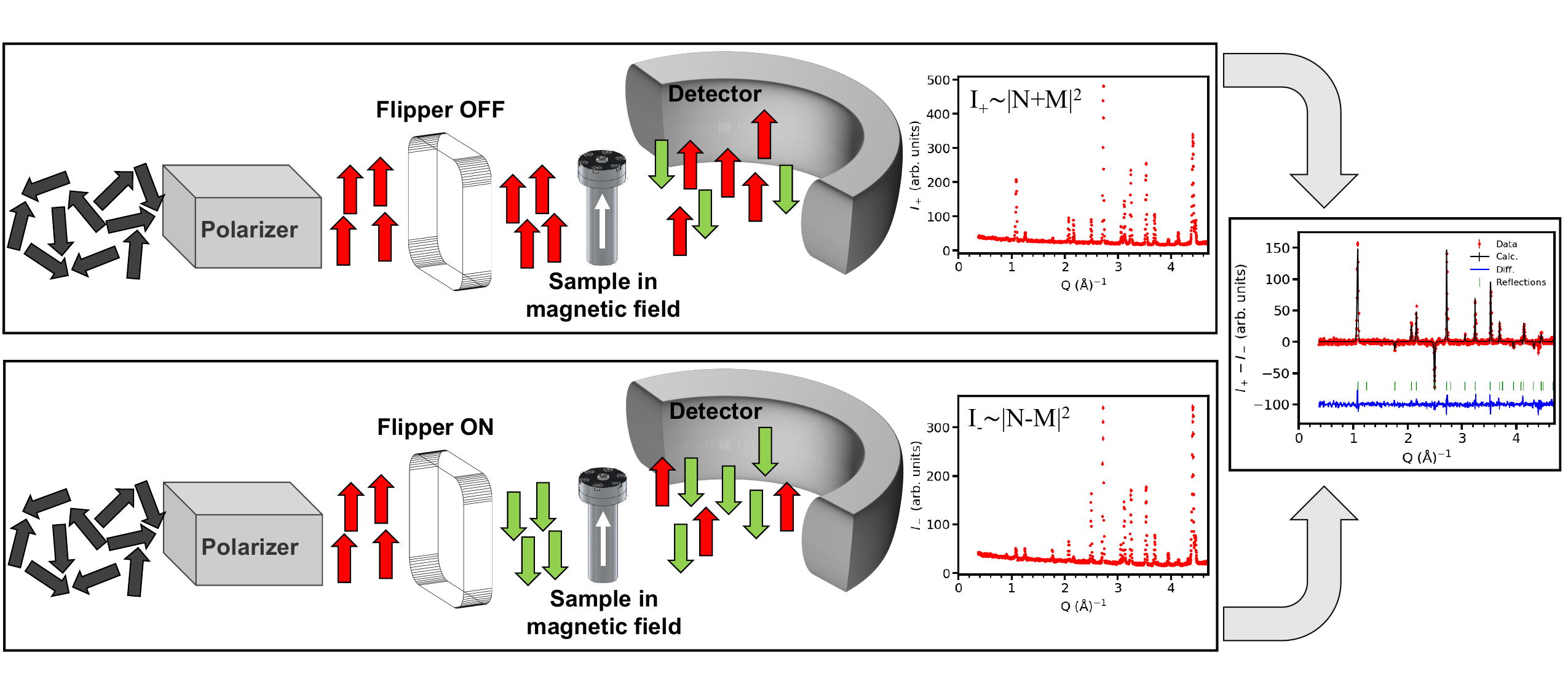} % Use .pdf or .eps if available
    \caption{\label{fig:Polarized_set_up} Schematic of the pNPD measurements showing the two required set-ups with flipper On ($I_-$) and Flipper Off ($I_+$). Unpolarized neutrons (black arrows) travel through a polarization filter. For HB-2A this is a V-cavity that polarizes, filters out, and transmits only one polarized state of the neutron beam. This is indicated by the red up arrows. There is then an option to keep the polarization state unchanged (Flipper OFF, $I_+$) or flip the neutron polarization to the opposite case (Flipper ON, $I_-$). The polarized beam then diffracts from the sample, which itself has a net moment, and the scattered neutrons are measured in the detector. Example data for each case is shown, with the difference being taken to plot $I_+ - I_-$.
    }
\end{figure*}
%%%%%%%%%%%%
% End Figure
%%%%%%%%%%%%

Determining the local susceptibility tensor allows insight into the anisotropy of magnetic ions in a material that drives the long- and short-range magnetic order. This is particularly important in materials with rare earth magnetic ions that have large spin-orbit coupling that creates enhanced anisotropy. Once determined, through consideration of the magnetic ion site symmetry and field direction response, the susceptibility tensors can be utilized to visualize the anisotropy as local magnetization ellipsoids around the magnetic ion within the crystal structure. Diagonalizing the susceptibility tensor gives the eigenvectors and eigenvalues of the principle magnetic directions $\chi_1$, $\chi_2$, $\chi_3$. These are shown in Fig.~\ref{fig:Ellipsoids}(a) and provide the magnetic axes and magnetic ellipsoids. These are treated analogously to atomic displacement parameters (ADPs), $U_{ij}$ or $B_{ij}$, and can be visualized in a similar fashion using crystallographic tools, such as VESTA \cite{VESTA}. This provides a straightforward categorization in terms of Ising, planar (XY) or Heisenberg anisotropy, with examples shown in Fig.~\ref{fig:Ellipsoids}. Moreover, the specific ellipsoid geometry within the crystal provides an indication of the magnetic moment direction. Trivially, the degree of anisotropy ($\chi'$) can be defined as $\sqrt((\chi_1 - \chi_{\rm m}) + (\chi_2 - \chi_{\rm m}) + (\chi_3 - \chi_{\rm m}))$, where $\chi_{\rm m}$ is the mean susceptibility $\chi_{\rm m}=(\chi_1 + \chi_2 + \chi_3)/3$.

The local site susceptibility tensor can be determined from measurements of a sample in an applied magnetic field with the incident neutron polarization state either parallel ($I_+$) or anti-parallel ($I_-$) to the sample field direction. To use neutron diffraction to extract $\chi_{ij}$ consider the scattered intensity during a measurement, given by:
\begin{eqnarray}
I \propto F_N^2 + {\vec F}_{M\perp}^2 +  {\vec P}\cdot(F_N^*{\vec F_{M\perp}} + F_N{\vec  F_{M\perp}^*}) + b_{\rm inc}
\label{Neutron_Unpol_and_Pol_Int}
\end{eqnarray}

where $F_N$ and ${\vec F}_M$ are the nuclear and magnetic structure factors, ${\vec P}$ is the polarization of the incident beam, and the symbol (*) denotes the complex conjugate. $b_{\rm inc}$ is the incoherent scattering. Note that we neglect a chiral term from these discussions since it is not relevant to the half-polarized technique. If the polarization state of the neutron beam is not controlled, ${\vec P}$=0, then the third term in the equation does not contribute to the scattering. The equation therefore reduces to the familiar unpolarized neutron powder diffraction case where the nuclear and magnetic scattering contribute separately to the scattering additively and both coherent and incoherent scattering are observed. If the polarization is non-zero (${\vec P}\neq$0) then the third term is accessed, which is the magnetic-nuclear interference term and gives different intensities for different neutron polarization states. 

To analyze the data the flipping difference method is used, with the intensity difference of spin-up ($I_+$) and spin-down ($I_-$) incident neutron beam polarized measurements given by:
\begin{eqnarray}
 I_+ - I_- \propto  2(F_N^\star \langle {\vec F}_{M,\perp}  \cdot  {\vec P} \rangle  + F_N \langle {\vec F}_{M,\perp}^\star  \cdot  {\vec P} \rangle  )  
\label{equation:Polarized_Diff} 
\end{eqnarray}

where the angular brackets account for the powder averaging. This relationship reveals that the difference signal is only observed in pNPD when there is both a magnetic and nuclear signal. The signal is enhanced, as expected, at larger magnetic reflections, and also, counter-intuitively, at {\it larger nuclear} reflections. Moreover, the incoherent scattering from the sample, and any background or sample environment scattering, is removed. This, therefore, makes pNPD measurements highly sensitive to weak signals and can access small moments ($<0.1 \rm \mu_B$) or samples containing large incoherent scattering, such as hydrogen \cite{PhysRevMaterials.6.124407, Calder:cam5003}.

In equation~\ref{equation:Polarized_Diff} the nuclear (atomic) scattering structure factor, $F_N$, can be readily determined from either complimentary unpolarized neutron scattering measurements or x-ray scattering measurements. For all pPND measurements on HB-2A a zero field unpolarized measurement is collected to allow for this structural refinement. Additionally, the polarization is known from the instrumental calibration. The magnetic structure factor, consequently, remains as the only variable to be determined. This is given by ${\vec F}_{M,\perp}(Q)=\sum_i {\vec m_i}f_m(Q) \exp(iQ.r_i)$, where the sum is over the unit cell and $f_m(Q)$ is the magnetic form factor. In unpolarized measurements ${\vec m_i}$ is considered as the local moment on the magnet site on atom $i$ that forms long-range or short-range magnetic order. In the local site susceptibility tensor approach ${\vec m_i}$ is the induced magnetic moment on atom $i$ by a magnetic field. The magnetic scattering can now be described as a structure factor tensor:
\begin{eqnarray}
 {\vec F}_{M,\perp}(Q)=\sum_i {\vec \chi_{ij}}f_m(Q) \exp(iQ.r_i)\cdot {\vec H}
\label{equation:Polarized_magn} 
\end{eqnarray}

Modeling the data is therefore reduced to a problem of determining the local site susceptibility at the specific applied magnetic field, through consideration of scattering intensity and allowed symmetry rotation and translations of the site $i$. 

An analysis of the sum ($I_+ + I_-$) should also be undertaken simultaneously to the difference to provide reliable normalized units and also obtain further constraints on the site susceptibility tensor. The sum has the form:
\begin{eqnarray}
 I_+ + I_- \propto  F_N + {\vec F_{M,\perp}} + b_{\rm inc},
\label{equation:Polarized_difference} 
\end{eqnarray}

and therefore recovers the unpolarized scattering of separate structural and magnetic scattering.

\subsubsection{pNPD experimental technique}

The application of this technique was initially only performed on single crystal samples, however has been extended to powder samples. This is described in detail in Refs.~\onlinecite{Gukasov_2010, PhysRevResearch.1.033100}. As a test case comparisons have been made between single crystal and powder data on rare earth pyrochlores and the information obtained has been shown to be equivalent \cite{PhysRevLett.103.056402, PhysRevResearch.1.033100}, making pNPD a powerful technique to obtain local anisotropic behavior. 

Here, we will consider measurements taken on the HB-2A powder diffractometer. Compared to the total scattering measurements discussed in Section \ref{mPDFsection} there are modifications to the instrumental set-up. A schematic of the set-up is shown in Fig.~\ref{fig:Polarized_set_up}. The initial unpolarized beam with randomly oriented neutron spins is polarized into one spin state. This is achieved on HB-2A using a V-cavity filter. This V-cavity has Fe/Si supermirrors arranged in a V shape that are contained in a strong magnetic field. The neutron spins are polarized with equal numbers of spin-up and spin-down orientations. The supermirrors have a spin-dependent transmission: one spin state is transmitted and one is reflected thereby being filtered out. Geometric constraints of the mirror angles limit the use of V-cavities to longer wavelengths. As such the 2.41 $\rm \AA$ wavelength is exclusively used on HB-2A for pNPD. A neutron flipper is then placed in the polarized beam, with the ability to either maintain the neutron spin state (flipper off, spin-up) or flip the neutron spin state (flipper on, spin-down). The sample is placed in a vertical field magnet to supply a field parallel to the neutron spin states. The intensity of the scattering from the sample is different when the neutron beam has a spin-up polarized state or a spin-down polarized state, see equation \ref{Neutron_Unpol_and_Pol_Int}, and this gives different scattered intensities. These different intensities, $I_+$ (flipper OFF) and $I_-$ (flipper ON) are subtracted and subsequently analyzed following equation \ref{equation:Polarized_difference}. For a complete analysis, both the difference ($I_+$ $-$ $I_-$) and the sum ($I_+$ $+$ $I_-$) are modeled and can provide further constraints on the final model, as described later when we consider material examples.  

\subsubsection{pNPD data analysis}

Rietveld refinement and other whole pattern fitting methods are routine for crystal structure, magnetic structure, and total scattering analysis for unpolarized neutron powder diffraction measurements. Until recently this has been lacking for pNPD and has contributed to the limited use of this technique. The open-source software CrysPy has filled this gap and is used for all the pNPD data analysis in the studies presented in section \ref{Results_Discussion} \cite{cryspy}. 

%%%%%%%%%%%%
% Begin Figure
%%%%%%%%%%%%
\begin{figure*}
    \centering
    \includegraphics[trim=0cm 1.0cm 0cm 1.0cm,clip=true,width=0.99,width=1.0\linewidth]{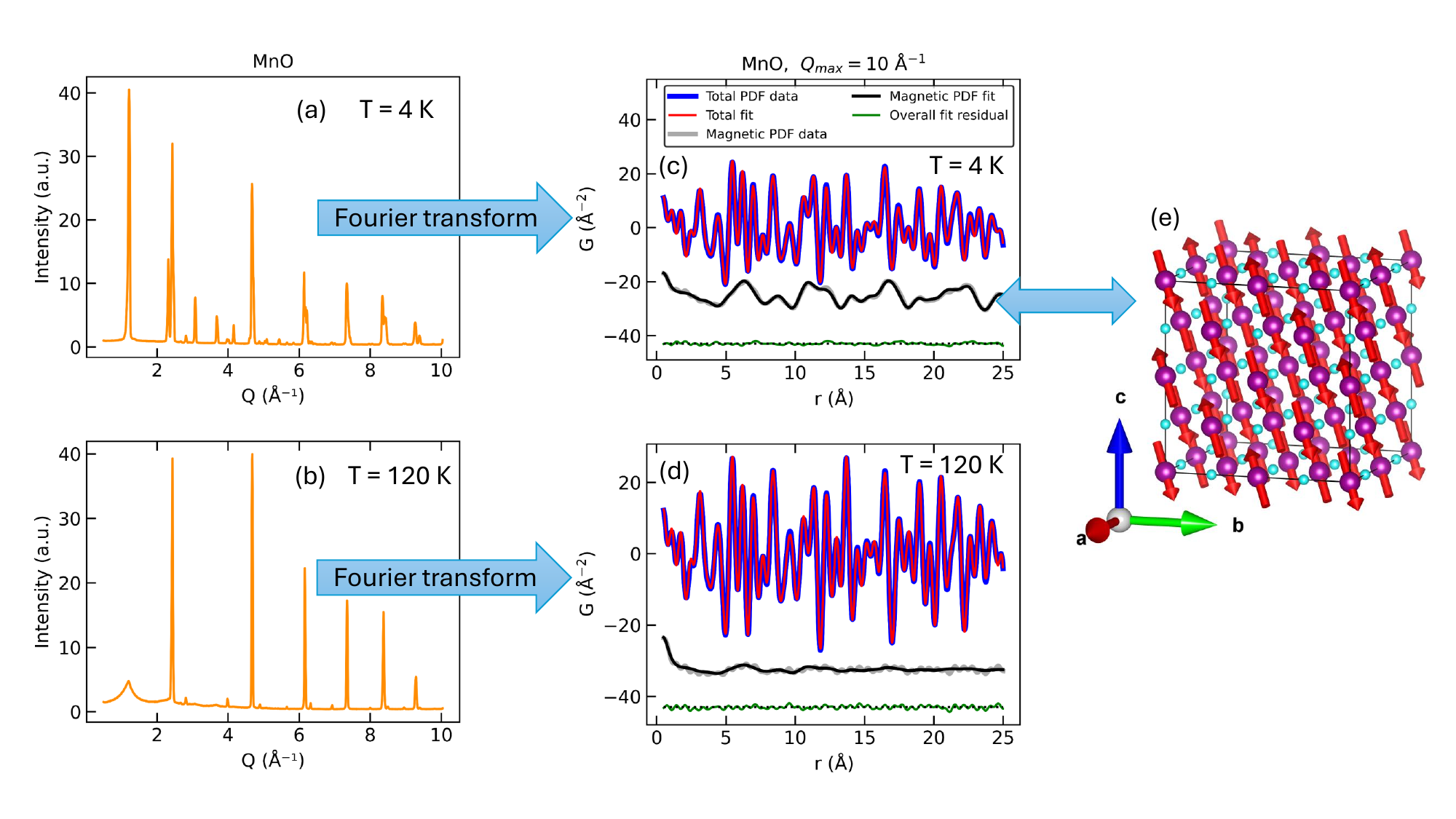} 
    \caption{\label{fig:MnO}(a, b) Total scattering structure function S(Q) for MnO at temperatures of 4 K and 120 K, respectively. (c) Atomic and magnetic PDF fits of MnO at 4 K using a fitting range of 0.5\,–\,25\,Å. The blue curve represents the total PDF data, while the red curve shows the total PDF fit. The gray, purple, and green curves correspond to the isolated magnetic PDF data, magnetic PDF fit, and overall fit residual, respectively. (d)  Atomic and magnetic PDF fits for MnO at 120 K. (e) Crystal and magnetic structure of MnO, with purple and cyan spheres representing Mn and O atoms, respectively.
    }
\end{figure*}
%%%%%%%%%%%%
% End Figure
%%%%%%%%%%%%

\subsection{Sample synthesis}
% \Dy215, \Ho227, MnO (purchased).

A high-purity (99\%) polycrystalline MnO powder sample was obtained from a commercial supplier (Alfa Aesar). A polycrystalline Dy$_2$TiO$_5$ sample was synthesized by heating stoichiometric amounts of Dy$_2$O$_3$ and TiO$_2$ through a solid-state reaction method. Dy$_2$O$_3$ was prefired at 1000$^\circ$C for 8 hours to remove moisture. The prefired Dy$_2$O$_3$ and TiO$_2$ powders were then mixed, pressed into a pellet, and fired at 1200$^\circ$C for 12 hours. After cooling, the pellet was reground, pressed again, and refired at 1500$^\circ$C for another 12 hours. The heating and cooling rates were maintained at 5$^\circ$C per minute. Tb$_2$Sn$_2$O$_7$ powder was prepared by a high-temperature, high-pressure hydrothermal method as described in Ref.~\cite{Danielle_Ce2Sn2O7}. A 50:50 ratio of SnO$_2$ to SnO (Alfa Aesar, 99.9\%) was mixed in a stoichiometric proportion with Tb$_4$O$_7$ (HEFA Rare Earth, 99.99\%) and added to fine silver (99.9\%) ampoules. Reduction of the mixed Tb(III)/Tb(IV) Tb$_4$O$_7$ to the Tb(III) state was performed in situ by the SnO reductant. A commercially available 50 wt\% CsOH solution (Acros, 99.9\%) served as the mineralizer. A nominal 1:2 ratio (i.e., 2 g powder and 4 mL mineralizer) was added to the silver ampoules, which were then weld-sealed. Ampoules were placed into a Tuttle-seal autoclave with deionized water added to attain a 100 MPa (15 ksi) counterpressure. Autoclaves were heated by ceramic band heaters to 700 ℃ over a period of two hours and held steady for three days before allowing to cool naturally in air over an eight-hour period. Recovered ampoules were vacuum filtrated with deionized water and dried with acetone to obtain a pure (\(>99\%\)) white powder. For the synthesis of Ho$_2$Ti$_2$O$_7$, dried Ho$_2$O$_3$ and TiO$_2$ powder with stoichiometric ratio was mixed and annealed in air at temperatures 1000, 1200, and 1450$^\circ$C with 20 hours for each.

\section{Results and Discussion}
\label{Results_Discussion}

\subsection{Model examples of mPDF and pNPD measurements on HB-2A}

\subsubsection{Magnetic PDF model example: MnO}

The first experimental demonstration of the mPDF technique was successfully performed a decade ago on MnO, which has a cubic rock salt structure at high temperature \cite{frandsen2015mPDFMnO}. The material undergoes a rhombohedral compression of the crystal lattice along the $\left\langle 111 \right\rangle$ direction below the N\'eel temperature of $T_N=118$ K. In MnO, the Mn$^{2+}$ spins exhibit a type-II antiferromagnetic structure described by the magnetic space group $C_c2/c$ ($\# 15.90$, BNS notation), where the spins align ferromagnetically within the (111) planes but antiferromagnetically between adjacent planes along the $\left\langle 111 \right\rangle$ direction and the spin alignment axis lies within the (111) plane maintaining a well defined in-plane orientation. MnO has become a de facto standard that has been used across instruments for evaluating the mPDF measurement and analysis. We follow this approach here with measurements and analysis on the HB-2A neutron powder instrument. 

The method of atomic and magnetic refinement in real space is used in this case with a Q$\rm_{max}$ of 10 \AA~ from the 1.12 $\rm \AA$ wavelength. The total scattering function $S(Q)$ (with the instrumental background subtracted) for MnO at 4 K is shown in Fig.~\ref{fig:MnO}(a), where magnetic Bragg peaks are observed at $Q =$ 1.22, 2.32, 3.1, 3.68, and 4.16 \AA$^{-1}$. The distinct and sharp magnetic peak at $Q=$ 1.22 \AA~ broadens out and merges into a diffuse feature above the ordering temperature at 120 K, as shown in Fig.~\ref{fig:MnO}(b). These $S(Q)$ patterns were Fourier transformed to generate the total PDF patterns in real space. These total PDF patterns include both atomic and magnetic scattering. To isolate the magnetic scattering pattern in real space, atomic PDF fits were performed using PDFgui \cite{Farrow2007PDFgui}. Once the atomic fit is complete, the fit residual (the difference between the PDF neutron data and the calculated atomic PDF fit) contains the magnetic PDF data along with any other artifacts present in the data. This residual of the atomic PDF fit is treated as the mPDF data in the subsequent mPDF modeling. Alternatively, simultaneous atomic and magnetic PDF fits can be performed using the Python-based software package \texttt{diffpy.mpdf}~\cite{frandsen2022diffpympdf}. For the calculations here the simultaneous atomic and magnetic fitting approach was used. Figure~\ref{fig:MnO}(c) shows the atomic and magnetic PDF fits of MnO at 4 K over a real space fitting range of 0.5 - 25 \AA. The blue and red curves represent the total PDF data and total calculated PDF fit, respectively. It should be noted that the total calculated PDF fit is the sum of atomic and magnetic fits. The gray and purple curves represent the isolated mPDF data ($d_{\mathrm{mag}}(r)$) and mPDF fit (offset for clarity). The overall fit residual is represented by the flat green curve. The fit shows very good agreement to the data, with $R_w = 5.3\%$ and the spin direction is consistent with the results of the previously reported literature \cite{frandsen2016verification,frandsen2022diffpympdf}. The sign of the peaks in the mPDF pattern reflects spin orientation: negative peaks signify antiparallel alignment, whereas positive peaks indicate parallel alignment. The local ordered moment determined from the mPDF fit is 4.8$\mu_B$ which is consistent with expectations for a magnetic S = 5/2 Mn$^{2+}$ ion.

Similar atomic and magnetic fits were also performed on the 120 K data,  corresponding to the short-range ordered paramagnetic state. Fig.~\ref{fig:MnO}(d) displays the atomic and magnetic PDF fits for MnO at 120 K. The mPDF $d_{\mathrm{mag}}(r)$ pattern at 120 K is reduced in magnitude as compared to 4 K, however, it clearly exhibits antiferromagnetic features in the low-\textit{r} region, as evidenced by broad negative and positive peaks. The fit quality is quantified by a goodness-of-fit value of $R_w = 4.7\%$. The isotropic correlation length at 120~K is determined to be $7.83 \pm 0.79$~\AA, obtained by applying an exponential decay envelope ($e^{-r/\xi}$) in the mPDF calculation, with $\xi$ used as a fitting parameter. The locally ordered moment is found to be $2.73 \pm 0.26 $~$\mu_B$.

\begin{figure*}
    \centering
    \includegraphics[width=0.89\linewidth]{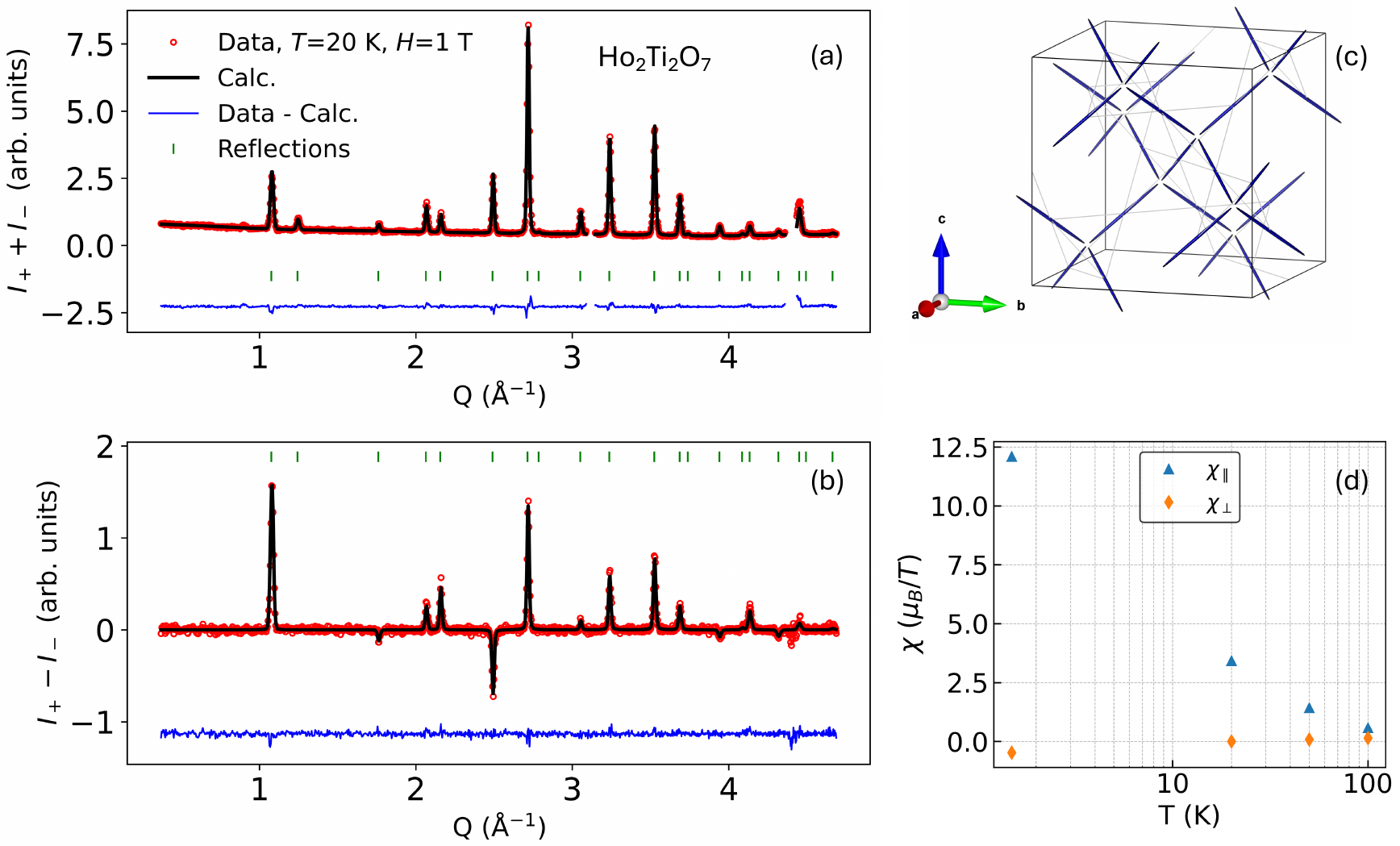} % Use .pdf or .eps if available
    \caption{\label{fig:Ho227_pol} The measured and calculated polarized neutron powder diffraction patterns of \Ho227 for the (a) sum and (b) difference at \textit{T}= 20 K, \textit{H}=1 T. (c) Magnetization ellipsoids for Ho are shown within a unit cell at \textit{T}= 20 K, \textit{H}=1 T. (d) Parallel ($\chi_{\parallel}$) and perpendicular ($\chi_{\perp}$) components of the susceptibility as a function of temperature of \Ho227 at field \textit{H}=1 T.}
\end{figure*}
%%%%%%%%%%%%
% End Figure
%%%%%%%%%%%%

\subsubsection{pNPD model example: \Ho227}

Having benchmarked the mPDF method using MnO, we now turn to pNPD measurements on Ho$_2$Ti$_2$O$_7$, which has become the de facto standard material for this technique. Ho$_2$Ti$_2$O$_7$ is a canonical geometrically frustrated material with fluctuating spins at all temperatures. The anisotropic single ion Ho$^{3+}$ behavior drives the short-range magnetic ordering.  Ho$_2$Ti$_2$O$_7$ offers a model example for pNPD measurements on HB-2A, with previous powder and single crystal studies being reported  \cite{PhysRevResearch.1.033100, PhysRevLett.103.056402}. 

The data was collected with spin up ($I_+$) and spin down ($I_-$) polarized incident neutron beam at 1.5 K, 20 K, 50 K, and 100 K all under an applied field of 1 T. Data for 20 K is shown in Fig.~\ref{fig:Ho227_pol}(a,b). The intensity changes with polarized state and this can be viewed by considering the flipping difference ($\Delta I$=$I_+ - I_-$) for 20 K in Fig.~\ref{fig:Ho227_pol}(b). 

To extract quantitative information the polarized data was refined using the CrysPy software to extract $\bar{\chi_i}$ for all temperatures \cite{PhysRevResearch.1.033100}. This is shown in Figs.~\ref{fig:Ho227_pol}(a,b) for 20 K. Inspection of the sum ($I_+ + I_-$) and difference ($I_+ - I_-$) data reveals that certain reflections show a clear difference in intensity for different neutron polarized states, while other reflections show no observable change. For example, consider the two peaks at 1.08 $\rm \AA^{-1}$ and 1.25 $\rm \AA^{-1}$, corresponding to the (111) and (200) reflections, respectively. Both reflections are enhanced in the sum data. Considering the difference data, however, only the (111) reflection shows a large intensity change, whereas for the (200) reflection there is no signal observed in the difference. This was utilized in the data analysis by refining both the sum ($I_+ + I_-$) and the difference ($I_+ - I_-$) and highlights the importance of doing this for all pNPD measurements. 

The local site susceptibility can be visualized as magnetization ellipsoids around the ion, which is shown in Fig.~\ref{fig:Ho227_pol}(c) for 20 K. At the lowest temperature measured of 1.5 K, this magnetization density shows highly anisotropic Ising behavior, with the magnetization density constrained to the local $\left\langle 111 \right\rangle$ directions. As the temperature increases the behavior becomes more isotropic. For the case of the Ho ion at the 16$d$ Wyckoff position in $Fd\bar{3}m$ space group the symmetry constraints lead to $\chi_{11}=\chi_{22}=\chi_{33}$ and $\chi_{12}=\chi_{13}=\chi_{23}$. Consequently, there are only two independent variables in the matrix tensor with the principle axes of the magnetization ellipsoids along the four local $\left\langle 111 \right\rangle$ directions. We follow the treatment in Refs.~\onlinecite{PhysRevLett.103.056402, PhysRevResearch.1.033100} to define $\chi_{\parallel}$ and $\chi_{\perp}$ terms as $\chi_{parallel}$=$\chi_{11}+2\chi_{12}$ and $\chi_{perp.}$=$\chi_{11}-\chi_{12}$ and plot these values for each temperature in Fig.~\ref{fig:Ho227_pol}(d). The $\chi_{\parallel}$ and $\chi_{\perp}$ behavior has been reported for several rare earth titanates from polarized measurements \cite{PhysRevLett.103.056402}. There is a strong temperature dependence of $\chi_{parallel}$/$\chi_{perp.}$, which indicates a change in the local anisotropy.

The results in Fig.~\ref{fig:Ho227_pol} for Ho$_2$Ti$_2$O$_7$ show the expected response and strong agreement with previous powder and single crystal measurements \cite{PhysRevResearch.1.033100, PhysRevLett.103.056402}. 

%%%%%%%%%%%%
% Begin Figure
%%%%%%%%%%%%
\begin{figure*}
    \centering
    \includegraphics[width=0.9\linewidth]{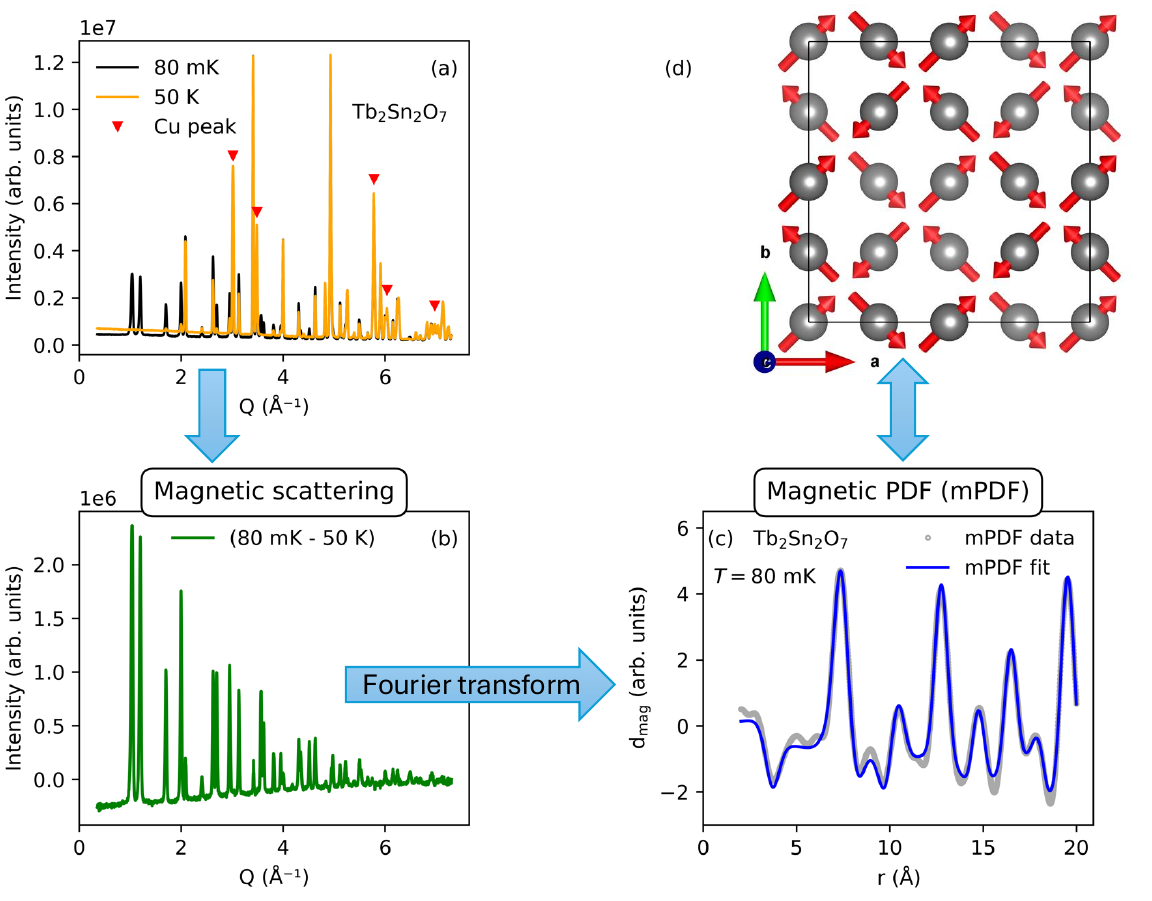} % Use .pdf or .eps if available
    \caption{\label{fig:TSO}(a) Total scattering structure function of Tb$_2$Sn$_2$O$_7$ at 80 mK and 50 K data, represented by black and orange curves, respectively. Red triangles represent the Cu peaks. (b) Intensity difference between the 80 mK and 50 K data. (c) mPDF data (grey symbols) for Tb$_2$Sn$_2$O$_7$ at 80 mK and the corresponding mPDF fit (blue line). (d) Magnetic structure of Tb$_2$Sn$_2$O$_7$.}
    
\end{figure*}
%%%%%%%%%%%%
% End Figure
%%%%%%%%%%%%

\subsection{\Tb227 investigated with mPDF}

Geometrically frustrated pyrochlores with rare-earth based ions in the form $R_2$Ti$_{2}$O$_7$ ($R=$ Tb, Ho, Dy,Er), exhibit unconventional magnetic behavior which typically involves regions of short range spin ordering. \Tb227~is such a geometrically frustrated magnet that crystallizes in the cubic pyrochlore structure (space group $Fd\bar{3}m$). The material undergoes long-range magnetic order at 0.87 K, but is characterized by regions of short range order due to the frustrated lattice topology \cite{PhysRevB.78.174416}.

Accessing the low-temperature behavior of materials has not been possible for the mPDF technique, in part due to the lack of suitable sample environments on traditional total scattering instruments. The HB-2A diffractometer routinely measures samples to millikelvin temperatures and has recently commissioned a multi-sample changer capability for dilution refrigerator (DR) experiments. This allows not only multiple samples to be measured but also allows for empty sample cans to be simultaneously loaded and measured, without the usual time-consuming burden associated with measuring separate DR samples. Due to the low temperatures involved the method of reciprocal space temperature subtraction was employed in this example. As such the need for high Q$\rm _{max}$ is reduced so a wavelength of 1.54 \AA~was used, which has the highest flux on HB-2A. We note the sample was measured in a Cu sample can, which aids thermalization of the powder sample, but adds Bragg peaks and further motivates the Q-space temperature subtraction approach. In addition, the sample can was charged with an overpressure of helium, which is now standard practice for any measurement below 1.5 K at HB-2A to further ensure that the sample reaches millikelvin temperatures.

Fig.~\ref{fig:TSO}(a) shows the neutron total scattering structure function of \Tb227 at 80 mK (black symbols) and 50 K (orange symbols). At 50 K, there is no observable diffuse scattering, instead the behavior follows a form factor drop-off, indicating this temperature is suitable to be used as a high-temperature uncorrelated state. The red triangular symbols indicate Cu peaks arising from the sample holder. The atomic structure undergoes minimal changes below 50 K down to 80 mK, with minimal changes to the peak positions. The scattering pattern collected at 50 K was used as a reference and subtracted from the 80 mK data to remove the nuclear contribution and additional Cu peaks that arise from the Cu sample holder also being removed.  The intensity difference between the data at 80 mK and 50 K is shown in Fig.~\ref{fig:TSO}(b), this scattering is treated as magnetic only. This magnetic scattering pattern was then Fourier transformed with a $Q_{\mathrm{max}}$ = 7 \An$^{-1}$ to generate the real-space mPDF patterns. Fig.~\ref{fig:TSO}(c) shows the mPDF data and fit of \Tb227 at 80 mK. The gray symbol represents the mPDF data and blue curve represents the mPDF fit. The refined magnetic structure is an ordered spin ice with spins following the "two-in,two-out" arrangement, see Fig.~\ref{fig:TSO}(c), and is consistent with previous reports on this material~\cite{Mirebeau_2005_Tb2Sn2O7_spins}. The first negative peak, which arises from the first nearest neighbor (NN) distance, is due to the net antiferromagnetic correlations. Similarly, the positive peak at the second NN distance represents the net ferromagnetic correlations and so on. This highlights the intuitive way to visualize the magnetic correlations directly in the real space offered by mPDF.

%trim is left, bottom, right, top
%%%%%%%%%%%%
% Begin Figure
%%%%%%%%%%%%
\begin{figure*}
    \centering
    \includegraphics[trim=2cm 2.5cm 1cm 2cm,clip=true, width=0.99\linewidth]{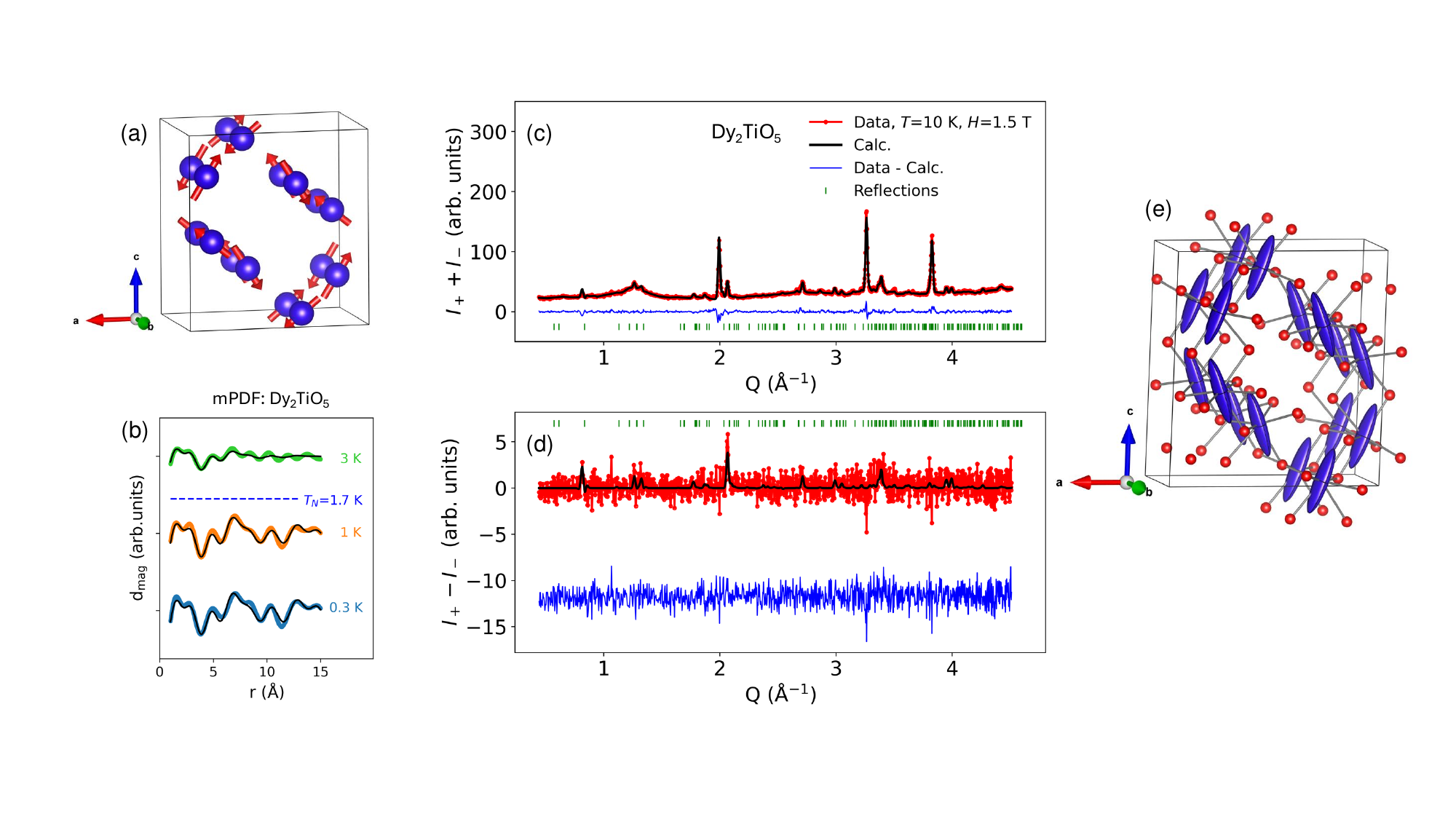} % Use .pdf or .eps if available
    \caption{\label{fig:DTO_pol} (a) Magnetic structure of \Dy215. mPDF pattern (offset for clarity) of \Dy215 at 0.3, 1, and 3 K, respectively. The measured and calculated polarized neutron powder diffraction patterns of \Dy215 for flipping (c) sum and (d) difference at \textit{T}= 10 K, \textit{H}=1.5 T. (e) Magnetization ellipsoids (blue) for Dy are shown within a unit cell. 
    }
\end{figure*}
%%%%%%%%%%%%
% End Figure
%%%%%%%%%%%%

\subsection{\Dy215 investigated with mPDF and pNPD}

As a final example, we present a combined mPDF and pNPD study on the frustrated rare-earth based material \Dy215. This material is related to the pyrochlore structure through the general formula $R_2$(Ti$_{2-x}$$R_x$)O$_{7-x/2}$ ($0 \leq x \leq 0.67$) that has been proposed for so-called stuffed pyrochlore-type solid solutions \cite{stuffed_pyrochlores_JSSC}. \Dy215 resides at the extreme end away from the cubic pyrochlores and forms a defect flourite structure \cite{magstructure_Dy}. \Dy215 has an orthorhombic structure (\textit{Pnma}) at room temperature and below. However, at higher temperatures, it transforms first into a hexagonal structure (\textit{P6\textsubscript{3}/mmc}) and subsequently into a cubic structure (\textit{Fm$\overline{3}$m}) \cite{park2018a2tio5,shepelev2006structures,aughterson2021synthesis}. The orthorhombic \Dy215 undergoes a 2D to 3D transition into a long-range ordered magnetic structure at 1.7 K \cite{magstructure_Dy}. The spin structure follows a coplanar model consisting of interwoven two-dimensional (2D) sheets extending along the [0 1 0] direction (see Fig.~\ref{fig:DTO_pol}(a)). There are two distinct Dy$^{3+}$ sites within the crystal structure.

\Dy215 presents unique challenges for mPDF and pNPD studies, principally the strong neutron absorption of Dy. Therefore, sample geometry and preparation have to be carefully considered. For the mPDF measurements, a vanadium can with annular geometry was used. The can had an overall diameter of 10 mm with the powder only occupying an annulus of 1 mm. Such a geometry mitigates the absorption from Dy, while still allowing for a suitable mass of powder. In the previous studies of  \Dy215 a flat-plate was utilized, however, this reduces the maximum $Q$ available significantly. For the pNPD measurements the loose powder grains are required to be fixed to stop any grain reorientation that would introduce texture when the magnetic field is applied. Typically several pellets or a single solid rod is made. However, this geometry was deemed too absorbing. Instead, for \Dy215 the powder was mixed with Si powder to dilute the Dy while still providing a large volume to use the large neutron beam. This powder mixture could then have been made into pellets. However, we chose an alternative route of loading the loose mixed powder of \Dy215 and Si into a standard vanadium can and then adding flourinert. Once the sample is cooled below room temperature the flourinert solidifies and will suspend the grains to prevent any reorientation when a magnetic field is applied. Flourinert does contribute to the diffraction pattern, most notably with a broad bump at 1.3 $\rm \AA^{-1}$, however, since the pNPD method takes a difference this is subtracted out since it will not change with polarization. See Figs~\ref{fig:DTO_pol}(c-d). Further challenges include the low ordering temperature of 1.7 K and the presence of two distinct Dy sites. As we have shown for mPDF the low ordering temperature enables a robust temperature subtraction and so is well suited to HB-2A. The presence of two ions adds a complication to the robustness of the data modeling due to additional variables in both mPDF and pNPD, however, is intrinsically not a fundamental limitation of these techniques.   

The pNPD measurements and analysis were performed within the linear M/H regime to satifsy the requirement for extracting the local site susceptibility tensor. The chosen temperature and field were 10 K and 1.5 T, which satisfy this requirement, as can be seen from the magnetization for \Dy215  in Ref.~\cite{magstructure_Dy}.  In the sum data ($I_+ + I_-$) the broad scattering is due to the flourinert. This could be modeled as background and subtracted out in the difference data ($I_+ - I_-$). Initially, modeling was performed on a zero-field unpolarized measurement at 10 K, where only the sum data contributes. This allowed the nuclear component to be refined in the absence of any magnetic scattering ($T_N$=1.7 K). Then the 1.5 T, 10 K data was modeled. Although the difference signal was observable even to a Q of above 3 \AA$^{-1}$, the signal was weak due to the dilution of the sample with Si and the addition of flourinert. There are two equivalent Dy sites. To aid modeling and reduce variables these were constrained to have the same magnitude in the presented results. Letting both Dy sites freely refine did not improve the quality of the fit. The refined site susceptibility tensor at 1.5 T was:

\begin{center} 
%\begin{eqnarray*}
$\begin{pmatrix}
		0.45(23) & 0 & 0.95(32)\\
			0& 1.02(24) & 0\\
			0.95(32) & 0 & 2.39(43)
	\end{pmatrix}$ $\mu_B$/Dy\\
%\end{eqnarray*}
\end{center}

The magnetization is shown as ellipsoids in the inset of Fig.~\ref{fig:DTO_pol}(c). This indicates anisotropic behavior with the magnetization primarily along the c-axis. Locally the anisotropy shows Ising spins that are along the Dy-O bond direction
% {\bf \color{red} [COMMENT: Could add a close up of this in the figure]}. 
This is consistent with the published magnetic structure \cite{magstructure_Dy}. For materials with no known magnetic structure, or those with only short range order, this pNPD analysis provides valuable details on the anisotropy and expected spin directions. The powder averaged magnetization can be extracted from the analysis and was found to be 1.7$\rm \mu_B$/Dy ion at 10 K and 1.5 T, in good agreement with the magnetization data in Ref.~\cite{magstructure_Dy}. It should be noted that this example was a challenging case for pNPD, however provides results consistent with the literature and as such strongly motivates further investigations of a wide variety of materials. 

The magnetic structure is now considered in both the ordered phase below $T_N=$1.7 K and in the short-range ordered phase at 3 K using mPDF. For the mPDF analysis, the total neutron scattering data was collected at 0.3 K, 1 K, 3 K, and 20 K. An approach similar to \Tb227 was used to obtain the mPDF $d_{\mathrm{mag}}(r)$ data on \Dy215. The high-temperature data at 20 K of \Dy215 was used as a reference temperature, which was subtracted from the low-temperature data to obtain the magnetic scattering patterns in \textit{Q}-space. These magnetic scattering patterns were subsequently Fourier transformed to generate the mPDF ($d_{\mathrm{mag}}(r)$) data in real space. Fig.~\ref{fig:DTO_pol}(b) represents the mPDF $d_{\mathrm{mag}}(r)$ data of \Dy215. The mPDF patterns are vertically offset for clarity to show the evolution with temperature through the transition. The mPDF model captures the mPDF data well at 0.3 K and 1 K, and the spin directions in these temperatures are consistent with the published magnetic structure \cite{magstructure_Dy}. Above the transition at 3 K, the mPDF pattern is reduced in magnitude in comparison to 0.3 K data, however strong short-range antiferromagnetic features are present and our mPDF model well captured the data to 10 \AA. Short range order was observed in previous measurements in Ref.~\cite{magstructure_Dy} on \Dy215, however detailed information was not able to be extracted. With the application of the mPDF methods described, detailed information is available on the short range spin structure above the magnetic transition, which can then be tracked as it evolves into the ordered phase.

\section{Conclusion}

The extension of capabilities on the HB-2A neutron powder diffractometer has been shown to be well-suited to materials with short range magnetic ordering. Specifically, the techniques of magnetic pair distribution function analysis and half-polarized neutron powder diffraction have been detailed. These distinct techniques require minimal instrument alterations and benefit from the addition of multi-sample changers, including those that can achieve under 100 mK temperatures, now available on HB-2A. The results obtained from either the mPDF or pNPD measurements can be utilized separately but together can be complementary to provide an overall understanding of the spin configuration in the material of interest. This is particularly applicable when considering materials with short range magnetic order. The materials best suited naturally encompass geometric frustrated rare-earth-based magnets but are not limited to these cases, with both techniques being applicable to any material with a magnetic ion.

\section*{acknowledgments}
This research used resources at the High Flux Isotope Reactor, a DOE Office of Science User Facility operated by the Oak Ridge National Laboratory. The beam time was allocated to HB-2A (POWDER) on proposal numbers IPTS-27680, IPTS-29447, and IPTS-32182. H.Z.~acknowledges support from the U.S. Department of Energy (DOE) under Grant No. DE-SC0020254. The Tb$_2$Sn$_2$O$_7$ crystals were synthesized at Clemson University as part of DOE BES grant DE-SC0020071. This manuscript has been authored by UT-Battelle, LLC under Contract No. DE-AC05-00OR22725 with the U.S. Department of Energy. The United States Government retains and the publisher, by accepting the article for publication, acknowledges that the United States Government retains a non-exclusive, paidup, irrevocable, world-wide license to publish or reproduce the published form of this manuscript, or allow others to do so, for United States Government purposes. The Department of Energy will provide public access to these results of federally sponsored research in accordance with the DOE Public Access Plan(http://energy.gov/downloads/doepublic-access-plan).

%\textbf{Author Contributions}
		
%\textbf{Declaration of Interests}
%The authors declare no competing interests.

% \section{References}
\bibliography{bibfile.bib}

\end{document}